\documentclass[]{spie}  %>>> use for US letter paper
\usepackage[]{graphicx}
\usepackage{tablefootnote}
\usepackage{amsmath}
\usepackage{epstopdf}

\newcommand{\refnum}[1]{Ref.~\citenum{#1}}
\def\mf{\mathbf}
\newcommand{\leftsub}[2]{{\vphantom{#2}}_{#1}{#2}}
\newcommand{\nCk}[2]{\ensuremath{\leftsub{#1}{C}_{#2}}}

\title{Gemini Planet Imager Observational Calibrations VIII: 
Characterization and Role of Satellite Spots} 

\author{Jason J. Wang\supit{a}, Abhijith Rajan\supit{b}, James R. Graham\supit{a}, Dmitry Savransky\supit{c}, Patrick J. Ingraham\supit{d}, Kimberly Ward-Duong\supit{b}, Jennifer Patience\supit{b}, Robert J. De Rosa\supit{b,e}, Joanna Bulger\supit{b}, Anand Sivaramakrishnan\supit{f}, Marshall D. Perrin\supit{f}, Sandrine J. Thomas\supit{g,h}, Naru Sadakuni\supit{i}, Alexandra Z. Greenbaum\supit{j,f}, Laurent Pueyo\supit{f}, Christian Marois\supit{k,l}, Ben R. Oppenheimer\supit{m}, Paul Kalas\supit{a}, Andrew Cardwell\supit{i}, Stephen Goodsell\supit{i}, Pascale Hibon\supit{i}, and Fredrik T. Rantakyr\"o\supit{i}, with the GPI team
\skiplinehalf
\supit{a} Department of Astronomy, UC Berkeley, Berkeley CA 94720, USA \\
\supit{b} School of Earth and Space Exploration, Arizona State University, Tempe, AZ 85287, USA; \\
\supit{c} Sibley School of Mechanical and Aerospace Engineering, Cornell University, Ithaca, NY 14853, USA; \\
\supit{d} Kavli Institute for Particle Astrophysics and Cosmology, Stanford University, Stanford, CA 94305, USA; \\
\supit{e} School of Physics, College of Engineering, Mathematics and Physical Sciences, University of Exeter, Exeter, EX4 4QL, UK; \\
\supit{f} Space Telescope Science Institute, 3700 San Martin Dr, Baltimore, MD 21218, USA; \\
\supit{g} NASA Ames Research Center,  Moffett Field, CA 94035, USA; \\
\supit{h} UARC, UC Santa Cruz, Santa Cruz CA 95064 USA; \\
\supit{i} Gemini Observatory, Casilla 603, La Serena, Chile; \\
\supit{j} Physics \& Astronomy Department, Johns Hopkins University, Baltimore MD, 21218, USA; \\
\supit{k} National Research Council of Canada Herzberg, Victoria, BC V9E 2E7, Canada; \\
\supit{l} University of Victoria, Victoria, BC, V8P 5C2, Canada; \\
\supit{m} American Museum of Natural History, New York, NY 10024, USA \\
}

\authorinfo{Further author information: (Send correspondence to J.J.W.)\\J.J.W.: E-mail: jwang@astro.berkeley.edu}
\begin{document} 
  \maketitle 

%%%%%%%%%%%%%%%%%%%%%%%%%%%%%%%%%%%%%%%%%%%%%%%%%%%%%%%%%%%%% 

\begin{abstract}
The Gemini Planet Imager (GPI) combines extreme adaptive optics, an integral field spectrograph, and a high performance coronagraph to directly image extrasolar planets in the near-infrared. Because the coronagraph blocks most of the light from the star, it prevents the properties of the host star from being measured directly. Instead, satellite spots, which are created by diffraction from a square grid in the pupil plane, can be used to locate the star and extract its spectrum. We describe the techniques implemented into the GPI Data Reduction Pipeline to measure the properties of the satellite spots and discuss the precision of the reconstructed astrometry and spectrophotometry of the occulted star. We find the astrometric precision of the satellite spots in an $H$-band datacube to be $0.05$ pixels and is best when individual satellite spots have a signal to noise ratio (SNR) of $> 20$. In regards to satellite spot spectrophotometry, we find that the total flux from the satellite spots is stable to $\sim 7$\% and scales linearly with central star brightness and that the shape of the satellite spot spectrum varies on the $2$\% level.
\end{abstract}

%>>>> Include a list of keywords after the abstract


%%%%%%%%%%%%%%%%%%%%%%%%%%%%%%%%%%%%%%%%%%%%%%%%%%%%%%%%%%%%%
\section{INTRODUCTION}
\label{sec:intro}  % \label{} allows reference to this section

The Gemini Planet Image (GPI) is a new facility-class adaptive optics (AO) instrument for the Gemini Observatory that is optimized for coronagraphic observations of bright ($I < 9$~mag.) natural guide stars. The principal science goal of GPI is imaging the environment of young ($\sim$ 125~Myr) stars in the solar neighborhood ($\sim$ 50~pc) to discover and characterize self-luminous planets and planetary debris disks. GPI uses low resolving power ($R \sim 45$) spectra to study exoplanet atmospheres and astrometry to quantify their kinematics \cite{2011PASP..123..692M}. 

Astrometry of exoplanets is necessary to discriminate between background objects and physically associated bodies and to measure their orbital elements. The semimajor axis, orbital inclination (relative to other components of the system), and eccentricity all carry important clues to their origin and evolution. Due to the large semimajor axis separation of GPI-detected planets, orbital periods are long (10---1000 yr) and orbital coverage is necessarily incomplete. While such fragmentary data can be used to reconstruct the orbit using modern statistical techniques (e.g. \refnum{2013ApJ...775...56K} and \refnum{2014arXiv1403.7520M}), the lack of a complete orbit places milli-acrsecond requirements on the accuracy and precision \cite{graham2007gemini}.

Near-IR ground-based spectrophotometry has to be corrected for strong wavelength dependent absorption by the terrestrial atmosphere (notably by H$_2$O and CO$_2$). Traditionally, this correction is applied using contemporaneous observations of calibration stars of known spectral type (typically A or G dwarf stars) at similar airmass. However, this requires additional planning and overhead for each star that is to be observed. It is more efficient when this calibration is done simultaneously with the science observations. 

\begin{figure}
   \begin{center}
   \includegraphics[height=6cm]{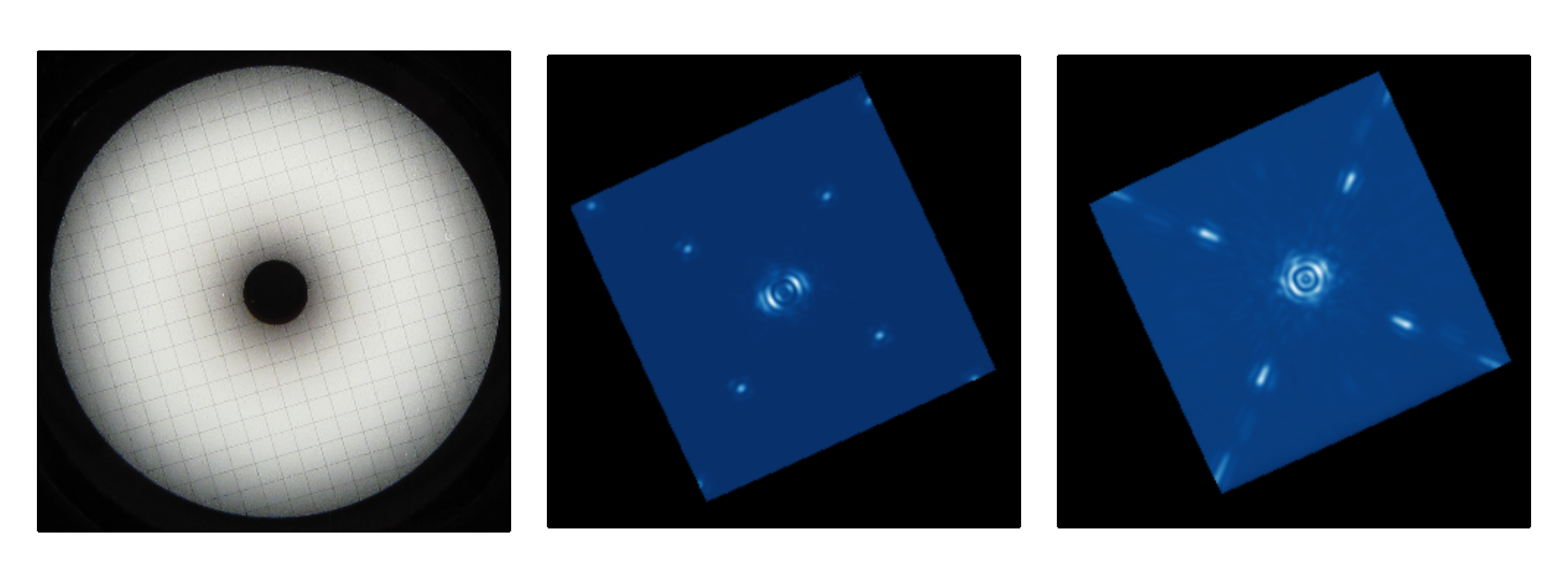}
   \end{center}
   \caption[example] 
%>>>> use \label inside caption to get Fig. number with \ref{}
   { \label{fig:grid_spots} 
{\it Left}: An image of the $H$-band apodized pupil mask fabricated by Aktiwave LLC, Rochester, NY.
{\it Center}: A wavelength slice from an $H$-band spectral datacube taken in the lab showing both the first order diffraction spots used for calibration and the second order spots that peek out from the edge of the detector. {\it Right}: A total intensity $H$-band image from a polarimetry mode datacube taken in the lab showing both the first-order and second-order diffraction spots. }
 \end{figure}

Coronagraphic observations pose a unique problem for differential spectrophotometry and astrometry of an exoplanet relative to its primary, which is occulted. To overcome this obstacle, GPI includes a square grid in a pupil plane which acts as a two-dimensional amplitude grating. The grid is superimposed on the the pre-occulter pupil apodizer (left image of Figure \ref{fig:grid_spots}). Diffraction of starlight from this grating injects first-order diffraction spots into the field of view for a given wavelength. In each wavelength channel in spectral mode, this creates reference spots that we term satellite spots (center image of Figure \ref{fig:grid_spots}). In broadband polarimetry mode, the satellite spots become streaks extending radially outward from the location of the star (right image of Figure \ref{fig:grid_spots}). These satellite spots preserve the information needed to reconstruct the spectrum and location of the occulted star: the diffraction pattern is centered on the true stellar position and is imprinted with an attenuated version of the stellar spectrum \cite{anand2006}\cite{marois2006}. In this report we discuss how we measure the satellite spots and investigate the spectrophotometric and astrometric properties of these diffraction spots. 

%-------------
%-------------

%%%%%%%%%%%%%%%%%%%%%%%%%%%%%%%%%%%%%%%%%%%%%%%%%%%%%%%%%%%%%
\section{Satellite Spot Extraction}
\label{sec:pipeline}

The raw data from the integral field unit is reduced using the GPI Data Reduction Pipeline (DRP)\cite{Perrin14}. For both spectral and polarimetry mode, the data are dark subtracted, corrected for flexure, and extracted into three-dimensional datacubes, where the $x$ and $y$ dimensions are angular coordinates on the sky. In spectral mode, the $z$-dimension of the datacube consists of the flux in 37 wavelength channels; in polarimetry mode, the $z$-dimension consists of the broadband flux in two orthogonal polarization modes.

There are a couple of steps that are important to properly measuring the satellite spots. Extracting the flux from each microlens (particularly in spectral mode) is challenging, especially in the presence of instrument flexure between the microlens array and the dectector \cite{Wolff14}. We characterize flexure shifts by taking argon arc lamp spectra after the telescope slews to a new target. Additionally, we apply a geometric distortion correction to all the images to rectify the position of the satellite spots \cite{Konopacky14}.

%%Note: spie standard is to refer explicitly to references by the ref number and NOT parenthetically, i.e., see Ref.~14.  I've added the \refnum command above to get the formatting right. Regular citations should just be superscripts as defined in the class.

%%-----------------------------------------------------------
\subsection{Spectral Mode} 
\label{sec:pipeline-spec}

\subsubsection{Locating \& Measuring the Satellite Spots}
\label{sec:pipeline-spec-spots}

Locating the satellite spots in the reduced spectral data cubes is typically 
achieved in two steps.  
First, an approximate set of spot locations is determined from one wavelength slice of the cube (this step can be bypassed if guesses for the locations of the satellite spots are manually supplied). These approximate locations are then scaled linearly with wavelength to find initial positions for all cube slices, which are then used to fit the precise location of each spot in each slice independently. 

The first step of the procedure is described in detail in \refnum{savransky2013computer}. Briefly, we assume that the four satellite spots form a perfect square in the image, and that they are relatively bright as compared with their neighboring pixels (although they need not be the brightest features in the image, nor even brighter than the median image value). These assumptions allow us to carry out an efficient search of the image. We iteratively construct a list of candidate spot locations, identifying the current brightest point in the image, and then removing it, along with a radius of $x$ pixels about it, from the search region (the value of $x$ is determined by the average size of a satellite spot in all cube wavelengths).  These bright images are stored in the expanding set $\{\mf r_i\}_{i=1}^N$, where each vector ($\mf r_i$) is the two-dimentional pixel coordinate of the candidate location.  As the list is being built, we conduct a breadth-first search for pixels sets forming perfect squares, where each node is a subtree whose branches represent all of the combinations of the root node with three other elements of  $\{\mf r_i\}$.  The value of all terminal nodes is binary, given by operating on the set of six distances between any four points ($\{\mf r_i\}_{i=1}^4$):
\begin{equation}\label{eq:cond4}
\{ d_k\}_{k=1}^6 = \left\{\Vert \mf r_i - \mf r_j \Vert : i,j \in \{\nCk{4}{2}\} \right\} \,,
\end{equation}
where $ \{\nCk{4}{2}\}$ is the set of two element tuples representing all of the combinations of four elements taken two at a time, without repetition. Assuming the set of distances is ordered by increasing magnitude, the four points define a square if and only if the first four distances ($\{ d_k\}_{k=1}^4$) are equal, the last two distances ($\{ d_k\}_{k=5}^6$) are equal, and the ratio of the magnitudes of the two subsets of distances is $\sqrt{2}$.

This search is made much more efficient by introducing a pruning heuristic, namely that all perfect squares contain right triangles. For any subset of three of the four candidate vertices, the corresponding subset of distances:
\begin{equation}\label{eq:cond3}
\{ d'_k\}_{k=1}^3 = \left\{\Vert \mf r_i - \mf r_j \Vert : i,j \in \{\nCk{3}{2}\} \right\} \,,
\end{equation}
must contain two elements that are equal, and one element that is $\sqrt{2}$ times larger.  This is a necessary, but not sufficient, condition for defining a square.  This condition allows us to prune whole branches of the tree, and greatly improve the search efficiency. This is also the smallest subtree that can be tested, as any two arbitrary points can form an edge of a square.

Once identified, the approximate satellite spot locations can be refined by fitting a template point spread function (PSF) to the area around the approximate location.  The usual approach is to assume a fixed-width Gaussian PSF as the template, and to use this in a matched filter (convolving with the pixels around the approximate location to find the best-fit offset). 

When extracting the peak flux of a satellite spot,  we first fit the full width half maximum (FWHM) to the actual data to calculate the standard devation of the Gaussian template. We then use this Gaussian template of fixed width in the convolution and fit for the height of the Gaussian.  This fitting step also allows us to relax the assumption of a symmetric PSF, as we can fit the FWHM in two orthogonal directions and thereby account for any PSF ellipticity.  

In order to ensure that our brightness assumption holds, we typically apply a high-pass filter to the image prior to beginning the satellite spot search. The high-pass filter is also applied when fitting for the precise spot locations (so as not to bias the astrometry with a background gradient) and can be optionally applied when fitting the spot peaks to exclude the residual AO halo from contrast calculations. We can also include specific constraints on the admissible size of elements in the distance sets in Eqs. (\ref{eq:cond4}) and (\ref{eq:cond3}), based on prior knowledge of the approximate relative separation of satellite spots in each wavelength band, which further allows us to prune the search tree and increase the efficiency of the original search.

\subsubsection{Locating the Central Star}
\label{sec:pipeline-spec-star}
In the current GPI pipeline, the central star is located by taking a simple mean of all $4 \times 37$ satellite spot positions. We rely on the fact that after distortion correction, opposite pairs of satellite spots are displaced equally and in the opposite direction in both $x$ and $y$ from the central star. For lab data, this technique works well as any single poorly measured satellite spot will have little effect on the measured central star position.

However, most on-sky data taken during GPI commissioning runs were without the atmospheric dispersion corrector (ADC) and suffers from atmospheric differential refraction (ADR)\cite{Hibon14}. Due to this, the central star is in a different position in each wavelength slice and a simple mean of all satellite spots can be biased in the case where a companion and its primary have different spectral shapes. The simplest approach to mitigate ADR effects is to measure the star position in each slice individually. This technique works well as long as the satellite spots are bright in each individual frame. We use this technique for all of the astrometric stability analysis in this paper to disentangle ADR effects from our satellite spot measurements. An illustrative description on how we locate the occulted star position is described in section \ref{sec:astrometry} and Figure \ref{fig:hip70931_sample}.

%%-----------------------------------------------------------
\subsection{Polarimetry Mode} 
\label{sec:pipeline-pol}
In broadband polarimetric images, the satellite spots are smeared radially outwards from the central star. The true location of the star would have all four streaks pointing towards it, excluding effects from ADR which will be discussed in section \ref{sec:astrometry-pol}. 

\subsubsection{Locating the Central Star}
\label{sec:pipeline-pol-star}

%-------------
   \begin{figure}
   \begin{center}
   \includegraphics[height=8cm, trim=0cm 2cm 0cm 2cm]{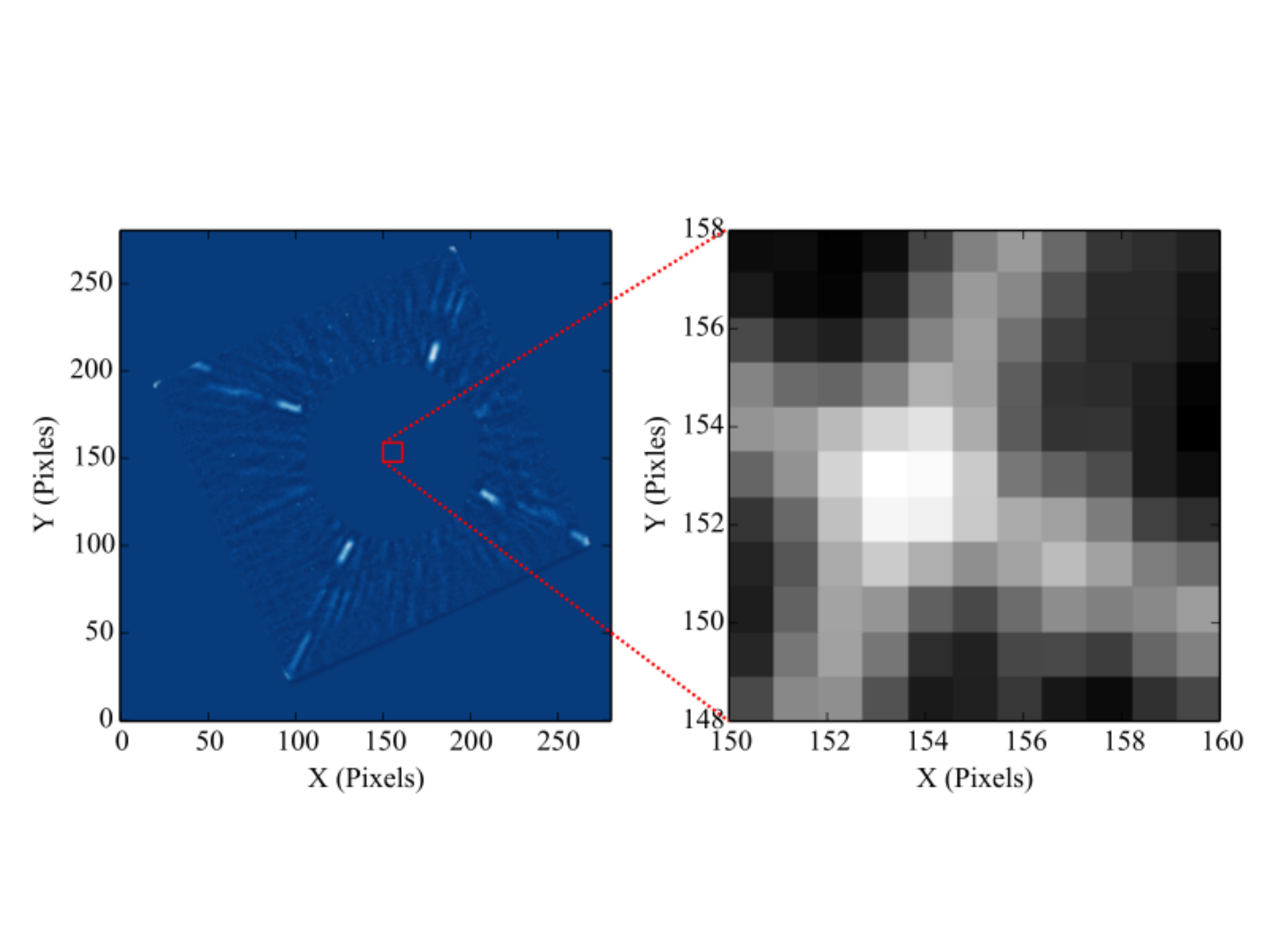}
   \end{center}
   \caption[example] 
%>>>> use \label inside caption to get Fig. number with \ref{}
   { \label{fig:radon} 
{\it Left}: A total intensity image of HIP~118666 in polarimetry mode with the central star masked. The image has been high-pass filtered using a median filter. The red box indicates the search area for the star center. Note that atmospheric differential refraction effects are apparent as the satellite spots do not all point back to the same location. 
{\it Right}: Map of the search box where each pixel represents the total intensity of all straight lines that pass through that pixel which we calculate using a Radon transform. This is before we interpolate this map to a finer resolution.}
   \end{figure} 
%-------------

To take advantage of the steaks of satellite spots, we take a simplified Radon transform of the image, implementing a technique similar to \refnum{Pueyo2014} with P1640 images. We will summarize the procedure here. First, we combine the two orthogonal polarization channels to form a total intensity image. We also run a high pass filter on the image by subtracting a median filtered image created with a $9 \times 9$ pixel box to calculate the median. If there are any visible objects (e.g. binaries) in the image, we manually mask them out at this point as they will interfere will the Radon transform. Then, we begin the Radon transform technique. We pick a guess for the center of the image and draw straight lines that go through this point. For each line, we compute the line integral of the image along that line (this corresponds to one point in Radon space). We then sum the square of the line integral along each of the lines (summing the square of each of the corresponding points in Radon space). Thus, we have gotten a measure of how much light is ``pointed" at this pixel. We do this for each pixel in a small search area around the initial guessed center of the image (see the left image of Figure \ref{fig:radon}). At this point, we have created a map of the intensity of the image in Radon space for each pixel (see right image of Figure \ref{fig:radon}). The pixel with the most power harbors the location of the central star. To achieve subpixel accuracy, we then interpolate this map to a finer resolution and pick out the subpixel that has the greatest value. 

Several of the steps we have outlined are introduced because of computational efficiency. Ideally, given infinite computational resources, we would interpolate the image before performing the Radon transform to achieve subpixel resolution. However, if we did do this, the computational time increases by a factor of order $N^4$ where $N$ is the over sampling factor (i.e. $N=2$ means doubling the image sampling). This is because the Radon transform method we use scales as $N^2$ and we need to do it for ever pixel in a search box which also grows as $N^2$. We also limit our search box to a small area near the true center for the same reason. Using the default $280 \times 280$ pixel image and an $11 \times 11$ pixel search box, this routine runs in a few seconds on a $3$ GHz core.

\subsubsection{Measuring Satellite Spot Fluxes}
Because the satellite spots are smeared into streaks, a precise flux measurement requires the development of new techniques to address this issue. We defer this work to a future paper and will not present any characterization of satellite spot photometry in polarimetry mode.

%%%%%%%%%%%%%%%%%%%%%%%%%%%%%%%%%%%%%%%%%%%%%%%%%%%%%%%%%%%%%
\section{Astrometric Stability}
\label{sec:astrometry}
In the following section, we present on the ability of the satellite spots to locate the occulted star. Because we can extract more information out of the satellite spots in spectral mode since they are not smeared out, we will focus most of our analysis on the spectral mode of GPI. We will then compare the spectral mode stability to the polarimetry mode stability. 

First, we note that it is difficult to establish the absolute accuracy and precision of the satellite spots. Images taken without the coronagraph that reveal both the star and the satellite spots would establish the reliability of reconstructing the star position. However, we have found it is impossible to achieve good signal to noise on the satellite spots without saturating the central star. Another technique of selecting long sequences of coronagraphic images and measuring the occulted star position is strongly affected by uncorrected tip/tip errors that cause the star to move and also cannot be used for this task. 

\begin{figure}
   \begin{center}
   \includegraphics[height=7cm]{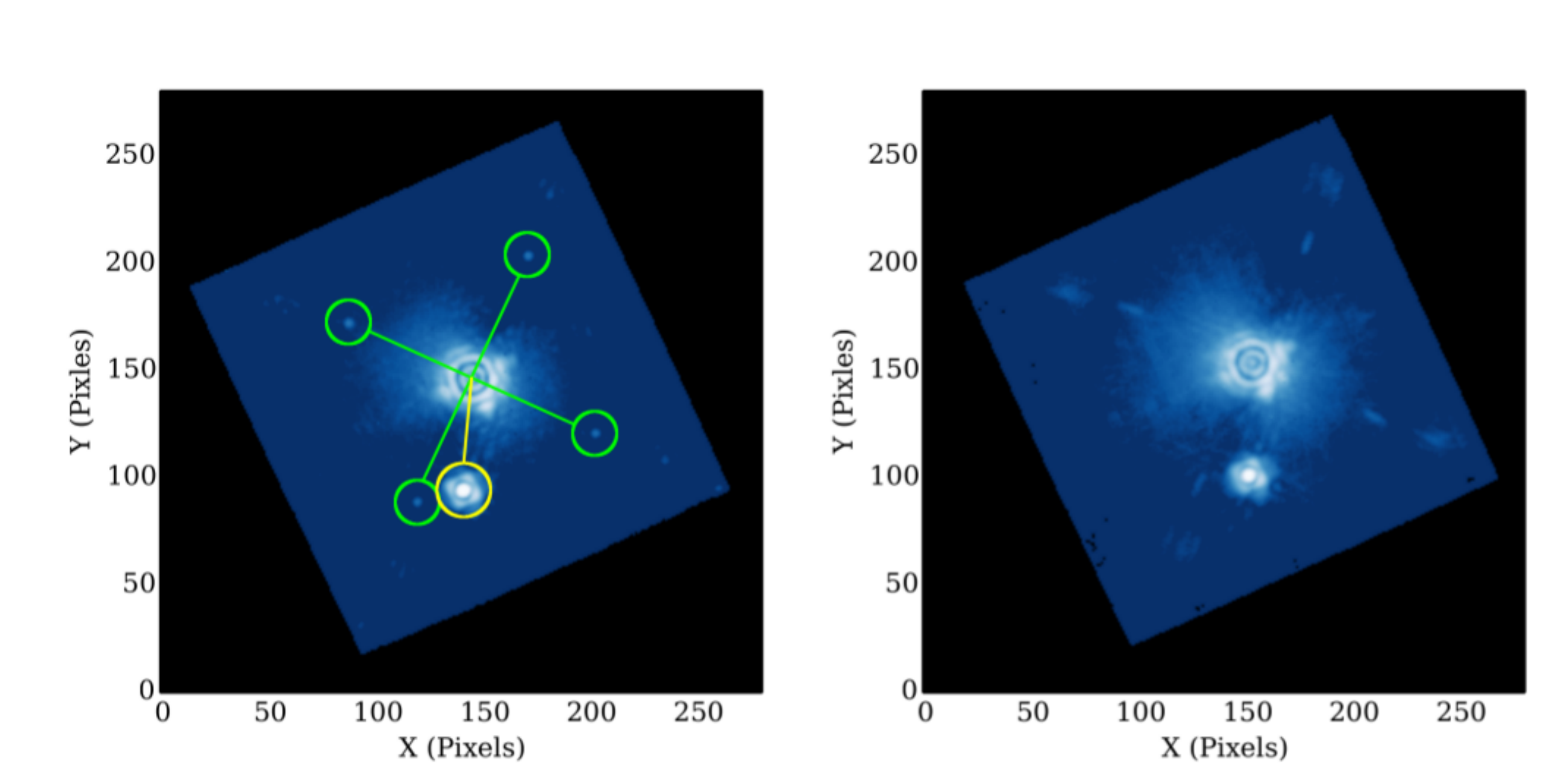}
   \end{center}
   \caption[example] 
%>>>> use \label inside caption to get Fig. number with \ref{}
   { \label{fig:hip70931_sample} 
Sample GPI images of the HIP~70931 binary system. These images have gone through the basic steps in the pipeline to be extracted into datacubes but no further processing of the data has been done. {\it Left}: The 1.663 $\mu$m wavelength slice of the first image taken in spectral mode on 2014 March 24 08:48 of HIP~70931. Overlaid on top is a schematic of how we measure the binary separation. The occulted star position is located using the satellite spots in green. And from there the binary separation can then be measured (yellow). {\it Right}: A total intensity $H$-band image from the first exposure of the 2014 March 24 polarimetry mode dataset of HIP~70931.
}
   \end{figure}

Using known astrometric binaries was a promising method we investigated. By measuring the relative separation of the binary companion and the occulted star, we could characterize the satellite spot astrometric accuracy. However, this method presents similar challenges. Known companions that fit into GPI's field of view are bright and saturate before we can obtain good signal to noise on the satellite spots. Thus, we have not been able to perform an absolute calibration of our ability to locate the occulted star. After an extensive search of all binaries, we found that the binary star system HIP~70931 has a faint enough companion and is therefore a suitable target to perform a binary separation analysis. Figure \ref{fig:hip70931_sample} shows some GPI data for this binary and a schematic of how we perform the astrometry analysis. The orbital parameters of this system have not yet been well characterized and cannot be used for an absolute calibration. We will instead present our relative stability analysis.

%-------------
%-------------

\subsection{Spectral Mode}
\label{sec:astrometry-spec}

In spectral mode, we measure the separation of the binary system in each wavelength slice of each datacube. Then, we average the separations calculated for each datacube to obtain the measured separation of that particular datacube. We do this to obtain similar centroiding precision as the regular pipeline processing step of finding the star center using all the wavelength slices together while allowing us to disentangle effects due to ADR. For HIP~70931, the companion and satellite spots are sufficiently bright to work with individual wavelength slices. The companion is also much brighter than the satellite spots ($\sim 10$ times brighter than the combined flux from all four satellite spots), so the uncertainty in the measured separation should probe errors in the satellite spot centroids. We use the same Gaussian matched filter approach for locating the companion as the satellite spots to reduce systematic biases.

HIP~70931 was observed on four separate nights in spectral mode (see Table \ref{tab:hip70931_sep} for a listing of nights and bands). We note that on 2014 May 10 and 2014 May 11, cloudy weather interfered with the observations and forced us to increase exposure times to compensate for extinction. Data on 2014 May 13 was taken during AO testing, causing the data quality to vary as AO parameters were adjusted. To mitigate this, we discarded any frames where the FWHM of the binary companion was greater than four pixels. We also discarded any frames in any of the datasets where the binary companion overlaps with a satellite spot.

%-------------
   \begin{figure}
   \begin{center}
   \includegraphics[height=9cm]{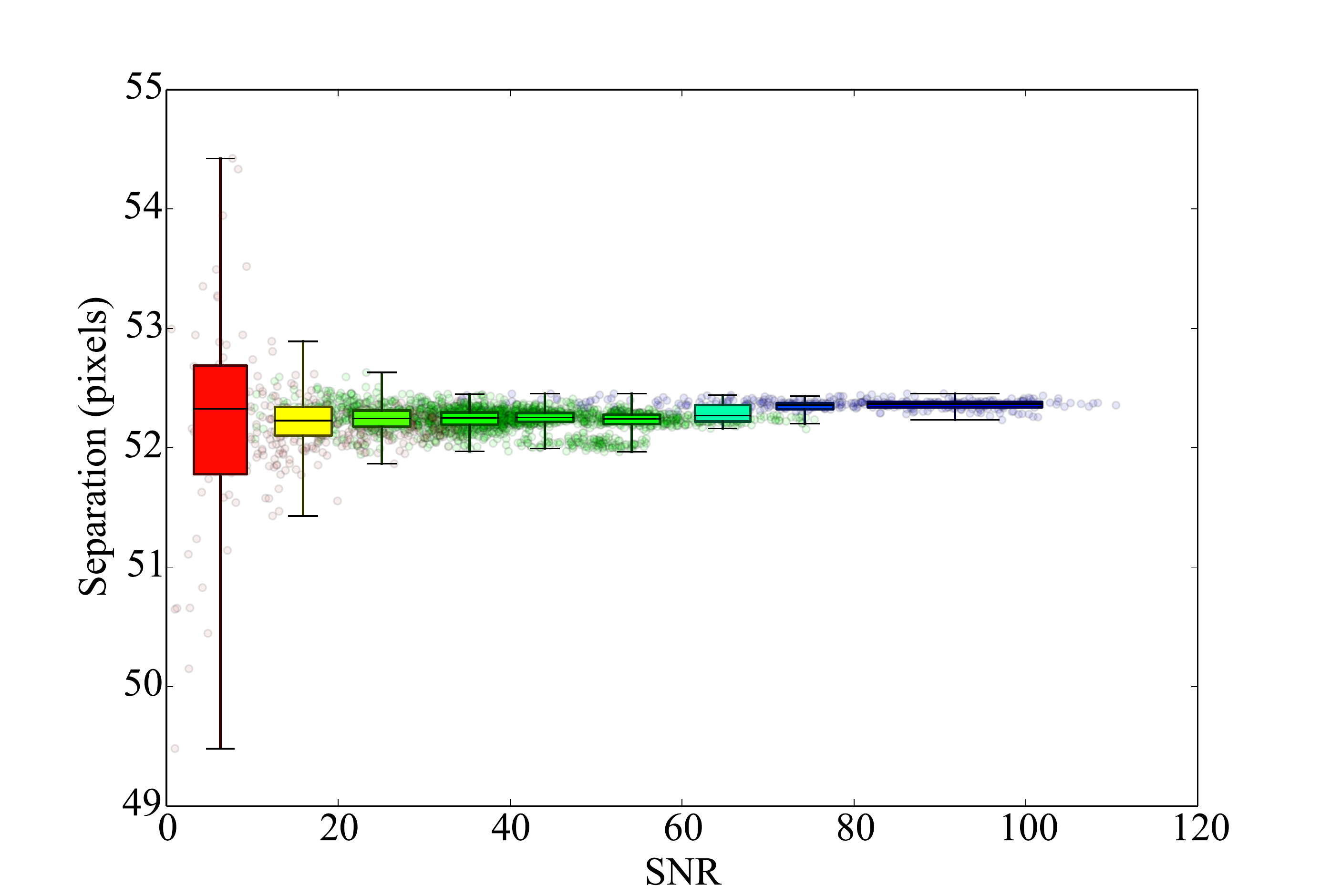}
   \end{center}
   \caption[example] 
%>>>> use \label inside caption to get Fig. number with \ref{}
   { \label{fig:hip70931_snr} 
Separation of the binary system HIP~70931 as a function of the average signal to noise ratio (SNR) of the four satellite spots in a wavelength slice of a datacube. Each faded circle in the background represents a separation calculated for an individual wavelength slice of a datacube, with color representing wavelength (red is $J$-band, green is $H$-band, and blue is {\it K1}, the first half of $K$-band). Overlaid on top are boxplots with the center of the box being the median, the edges of the box being the first and third quartiles of the data, and the edges of the whiskers being the most extreme points. Each boxplot represents a SNR range of 10 except for the last box which represents SNR $> 80$ and is larger in width. The color of the box indicates the fraction of points in each bin from each wavelength band (i.e. the red bin has almost exclusively points from {\it K1} and the yellow bin is a mix of $H$ and {\it K1} data). }
   \end{figure} 
%-------------

In Figure \ref{fig:hip70931_snr}, we plot the measured separation in each wavelength slice of each datacube as a function of average satellite spot signal to noise ratio (SNR). SNR was calculated by aperture photometry using an aperture radius of 5 pixels to measure the flux, a sky annulus between 8 and 12 pixels to measure the sky background, and the assumption that the scatter is Poisson noise. We see that the astrometric precision of the satellite spots decreases significantly when the average SNR of the four satellite spots in a single wavelength slice falls below $\sim 20$. In the following analysis of the astrometric precision of the satellite spots, we do not consider any individual wavelength slice where the average satellite spot SNR $< 20$. This SNR corresponds to a photon noise accuracy limit of $< 0.2$ pixels as given by FWHM/SNR for an individual satellite spot in $H$-band (typical FWHM of $< 4.0$ pixels). If we combine the four satellite spots, this would then limit the uncertainty on the position of the occulted star to $< 0.1$ pixels. This leads us to conclude that future observations taken with GPI should ensure SNR of individual satellite spots to be above 20 to ensure the central star can be located precisely.

Within a single exposure, the errors seem to be dominated by photon noise. In Table \ref{tab:hip70931_sep}, we log the average scatter in the separation measured in each slice of a datacube and the theoretical limit on centroiding accuracy due to purely photon noise for a single slice. We calculate this limit using the expression $($FWHM/fSNR$)/2$. fSNR is the average signal to noise of the four satellite spots after running a high-pass filter (median filter subtraction with a $9 \times 9$ pixel box) on the image, since this is done before locating the spots in our algorithm. We divide the centroid precision of a single satellite spot by $\sqrt{4}=2$ to get the precision of locating the occulted star since we take the mean of the four satellite spots to locate the star. We find that the scatter in the measured separation approaches the photon noise limit of the central star position. Thus, within a single datacube we conclude that the majority of the error in centroiding on the central star is dominated by photon noise in the satellite spots.

%%%% TODO: Make sure footnote is in the right page when we submit
\begin{table}
\caption{Measurement of the HIP~70931 binary separation for various observation sequences in spectral mode. We list the average internal scatter of measurements from wavelength slices within a single datacube (``Internal Scatter"), the photon noise limit for centroiding precision of occulted star for an individual slice ("$($FWHM/fSNR$)/2$"), and the mean separation and sample standard deviation as calculated from combining datacubes in an observing sequence (``Measured Separation").  } 
\label{tab:hip70931_sep}
\begin{center}       
\begin{tabular}{|c|c|c|c|c|c|} %% this creates two columns
% |l|l| to left justify each column entry
% |c|c| to center each column entry
% use of \rule[]{}{} below opens up each row
\hline
\rule[-1ex]{0pt}{3.5ex}  \shortstack{Observation Start \\ (UT)} & {Band} & {Frames} & \shortstack[c]{Internal Scatter\protect\footnotemark  \\ (pixels)} & \shortstack[c]{$($FWHM/fSNR$)/2$ \\ (pixels)}  & \shortstack[c]{Measured Separation\protect\footnotemark  \\ (pixels)}  \\
\hline
\rule[-1ex]{0pt}{3.5ex}  2014 March 24 08:48 & $H$ & 7 & $0.03$ & 0.03 & $52.26 \pm 0.01$\\
\hline
\rule[-1ex]{0pt}{3.5ex}  2014 March 24 09:07 & $H$ & 4 & $0.03$ & 0.03 & $52.04 \pm 0.02$\\
\hline
\rule[-1ex]{0pt}{3.5ex}  2014 May 10 03:52 & $H$ & 10 & $0.04$ & 0.03 & $52.26 \pm 0.05$\\
\hline
\rule[-1ex]{0pt}{3.5ex}  2014 May 10 05:46 & $J$ & 9 & $0.03$ & 0.02 & $52.36 \pm 0.03$\\
\hline
\rule[-1ex]{0pt}{3.5ex}  2014 May 11 04:22 & $H$ & 6 & $0.08$ & 0.06 & $52.26 \pm 0.04$\\
\hline
\rule[-1ex]{0pt}{3.5ex}  2014 May 11 05:08 & {\it K1} & 7 & $0.10$ & 0.06 & $52.17 \pm 0.03$\\
\hline
\rule[-1ex]{0pt}{3.5ex}  2014 May 13 03:28 & $H$ & 31 & $0.04$ & 0.04 & $52.25 \pm 0.05$\\
\hline
\rule[-1ex]{0pt}{3.5ex}  All $H$ Combined & $H$ & 58 & 0.05 & 0.04 & $52.24 \pm 0.07$\\
\hline
\rule[-1ex]{0pt}{3.5ex}  Clipped $H$ Combined\protect\footnotemark& $H$ & 54 & 0.05 & 0.04 & $52.26 \pm 0.05$\\
\hline
\end{tabular}
\end{center}
\end{table}

\begin{figure}
   \begin{center}
   \includegraphics[height=7.2cm]{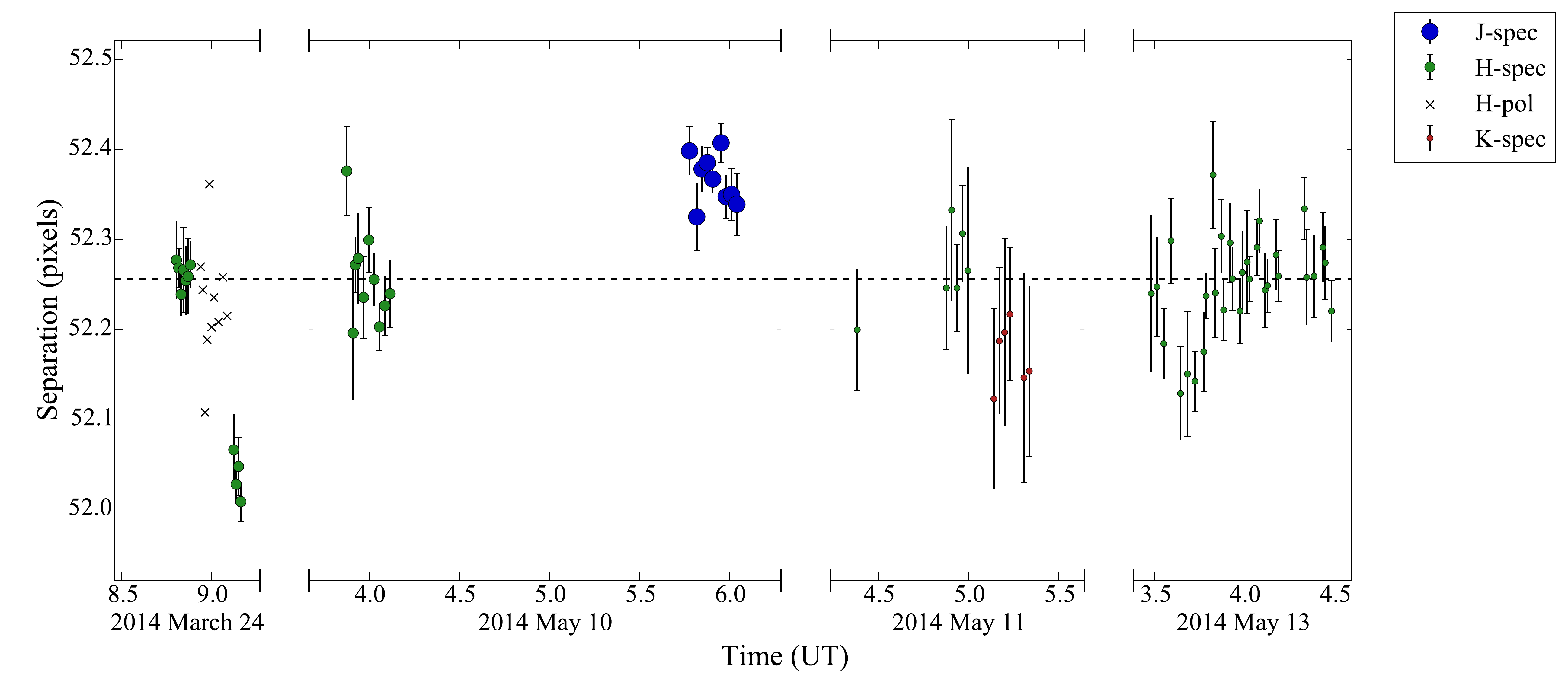}
   \end{center}
   \caption[example] 
%>>>> use \label inside caption to get Fig. number with \ref{}
   { \label{fig:hip70931_time} 
Measured separation of the binary system HIP~70931 as a function of time. Each circle represents the mean separation measured in a single datacube, with error bars representing the rms scatter within the datacube and color representing wavelength of the data as denoted in the legend. The three different sizes of markers represent 3 different signal to noise bins: the smallest symbols indicate datacubes with a median satellite spot SNR of less than 40, medium sized ones are for SNR between 40 and 60, and the largest markers for SNR greater than 60. Black crosses represent separations calculated in polarimetry mode in $H$-band. The horizontal dashed line represents the mean separation calculated with all of the $H$-band spectral data, excluding the four images with abnormally low separation taken 2014 March 24. }
   \end{figure}

We can no longer reach down to the photon noise limit when averaging datacubes together. When looking at the measured separation between datacubes in a single epoch, we would expect the photon noise floor for the sample standard deviation to be around $($FWHM/fSNR$)/\sqrt{37\times4} \approx 0.005$ pixels in $H$-band since we are now averaging over all satellite spots in all slices of a datacube. The last column of values in Table \ref{tab:hip70931_sep} shows that we do not get down to that level of precision, hitting a noise floor between $0.01$ and $0.05$ pixels. In fact, some observing sequences saw an increase in scatter when combining values from different datacubes. Looking at Figure \ref{fig:hip70931_time}, we see trends in the measured separation as a function of time. It is especially pronounced in the last four exposures taken on 2014 March 24 and in the dip in the measured separation on 2014 May 13. 

We speculate this could be due to one or more of the following effects. Qualitatively from looking at the images as they evolve in time, we noticed that the PSF changes due to uncorrected wavefront aberrations. This could cause biases in our centroiding routine. In particular, the satellite spots, which are a diffraction phenomenon, do not have the same PSF as a star imaged on the detector. The satellite spots are slightly elongated along one axis due to diffraction since the wavelength channels span a finite wavelength range. Additionally, speckle noise near the satellite spots and diffraction spikes from the occulted star, which are aligned with the satellite spots, vary in time and can bias the centroiding if positioned correctly. However, we have not determined if any of these effects is the cause of the variations we see. We do note that we do not see any trends in the measured separation as a function of wavelength or as a function of the satellite spot distance from the occulted star (a proxy for plate scale).

\addtocounter{footnote}{-2}
\footnotetext{rms scatter of separations measured between slices of each individual datacube}
\addtocounter{footnote}{1}
\footnotetext{error quoted is sample standard deviation in the measured separation among datacubes}
\addtocounter{footnote}{1}
\footnotetext{removed data from the four exposures taken at 2014 March 24 09:07 that appear to be outliers (see Figure \ref{fig:hip70931_time})}

%-------------
%-------------

Altogether, in the four different epochs of observation in $H$-band plotted in Figure \ref{fig:hip70931_time}, we find that despite the uncertainties, the data agree very well. The mean separations listed in Table \ref{tab:hip70931_sep} are almost identical for these four epochs. When we combine these four nights of observations, we find the uncertainty in the occulted star position is $0.07$ pixels. If we eliminate the four outlier points from 09:00 on 2014 March 24, the uncertainty decreases down to $0.05$ pixels (or $\sim 0.7$ mas), which we take as the uncertainty in the central star position for an $H$-band datacube along one axis. When moving to other bands ($J$ and {\it K1}), we see a systematic disagreement with $H$-band on the order of $0.1$ pixels. Since almost all of the astrometric calibration and analysis has been done at $H$-band, it is not surprising that we have not accounted for some chromatic effects.

Our uncertainties in satellite spot astrometry should not greatly impact GPI's science performance though. This error term is below the $1.8$ mas per epoch requirement for GPI astrometry. Astrometric uncertainties in plate scale and distortion are of comparable magnitude\cite{Konopacky14}. Additionally, when it comes to GPI's main science goal, exoplanets, we will likely be limited by the low SNR of the exoplanet itself - we would need a SNR of greater than 30 for centroid errors due to photon noise from the exoplanet to fall under 0.1 pixels in $H$-band.

\subsection{Polarimetry Mode}
\label{sec:astrometry-pol}
Limited data are available to characterize satellite spot astrometry in polarimetry mode. We will focus on ten observations of the binary HIP~70931 taken on 24 March 2014 in between the two spectral mode observations discussed in the previous section. Using the Radon transform technique described in section \ref{sec:pipeline-pol-star} to find the occulted star, we measure a binary separation of $52.23 \pm 0.06$ pixels. The measured separation and error is comparable to the spectral data (see Figure \ref{fig:hip70931_time}).

We note the lack of systematic bias between the polarimetry and spectral data. All of the HIP~70931 data taken on 24 March 2014 are without the atmospheric dispersion corrector (ADC) at an airmass of $\sim 1.15$. In spectral mode, we were able to mitigate chromatic effects by analyzing each wavelength slice independently, something we are not able to do in polarimetry mode. Despite that, finding comparable separations indicate that atmospheric differential refraction (ADR) does not significantly affect the satellite spots astrometric precision in polarimetry mode, at least not for binary star systems taken in $H$-band at airmasses less than about $1.15$. It remains to be seen how great of an effect ADR is for planets, which differ more in color with respect to their host stars, and at low elevations when ADR effects are the largest. This however should be mitigated with the commissioning of the ADC \cite{Hibon14}. 

\begin{table}[!h]
\centering
\caption{Measurements of the grid ratio (ratio of brightness of the occulted star to the total brightness of all four satellite spots) and corresponding magnitude difference}
\begin{tabular}{|c|c|c|c|c|}
\hline
 Target	& Date	& Filter & Grid Ratio	& $\Delta m$ (mags)  \\
\hline 
$\beta$ Pic	& 2013-11-18	& $H$ & 5433 $\pm$ 1790	& 9.3 $\pm$ 0.4 \\
\hline 
$\beta$ Pic	& 2013-12-10	& $H$ & 5395 $\pm$ 953	& 9.3 $\pm$ 0.2 \\
\hline 
$\beta$ Pic	& 2013-12-11	& $H$ & 4569 $\pm$ 1936	& 9.2 $\pm$ 0.5 \\
\hline 
HD 118335	& 2014-03-25	& $H$ & 6147 $\pm$ 964	& 9.5 $\pm$ 0.2 \\
\hline
\end{tabular}
\label{tab:grid_ratio}
\end{table}

%%%%%%%%%%%%%%%%%%%%%%%%%%%%%%%%%%%%%%%%%%%%%%%%%%%%%%%%%%%%%%
\section{Spectrophotometric Stability}
\label{sec:phot}

\begin{figure}
\begin{center}
\includegraphics[width=12cm]{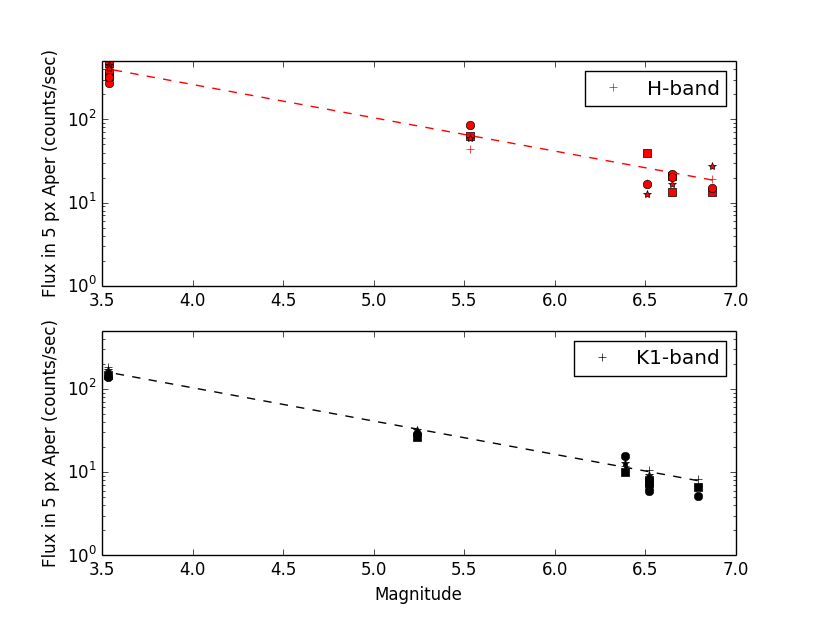}
\end{center}
\caption{\label{fig:th3} The average flux measured within the 11th to 14th wavelength slice of the reduced data cubes for the four satellite spots (differentiated by symbol) in the $H$-band (red dashed line and symbols), and $K1$-band (black dashed line and symbols). The fits to the measured fluxes as a function of magnitude are consistent with the expected relation of $m=-2.5\log(f)+c$.}
\end{figure}

We used data collected on targets observed in the coronagraphic mode, during the first two commissioning runs in November and December of 2013, in order to measure the spot flux as a function of target magnitude (Figure~\ref{fig:th3}). The observations were reduced using the typical reduction primitives within the GPI DRP. The background within each image was subtracted by applying a high-pass filter, with the satellite spots being masked to prevent self-subtraction. Aperture photometry was then performed on the four satellite spots within each wavelength slice of the reduced data cube. These measurements, obtained with the $H$ and $K1$ filters, show a linear trend of decreasing satellite spot flux as a function of stellar flux.

The ratio of the flux of satellite spots, where we average the flux of all four spots, to the flux of the star behind the occulting mask was also estimated using two different techniques and is presented in Table~\ref{tab:grid_ratio}. A detailed discussion of all the photometry results from the GPI photometric calibration program presented at this conference\cite{Maire14}.

\subsection{Long duration monitoring}
To measure the spectrophotometric stability of GPI, data taken on multiple stellar targets with time-series durations spanning several minutes to multiple nights were analyzed. The data presented in this paper were collected primarily in the first and second calibration runs over Nov, Dec 2013 and Mar 2014. 
%-------------
   \begin{figure}[!t]
   \begin{center}
   \includegraphics[height=13cm,trim=3cm 0.5cm 3cm 0cm]{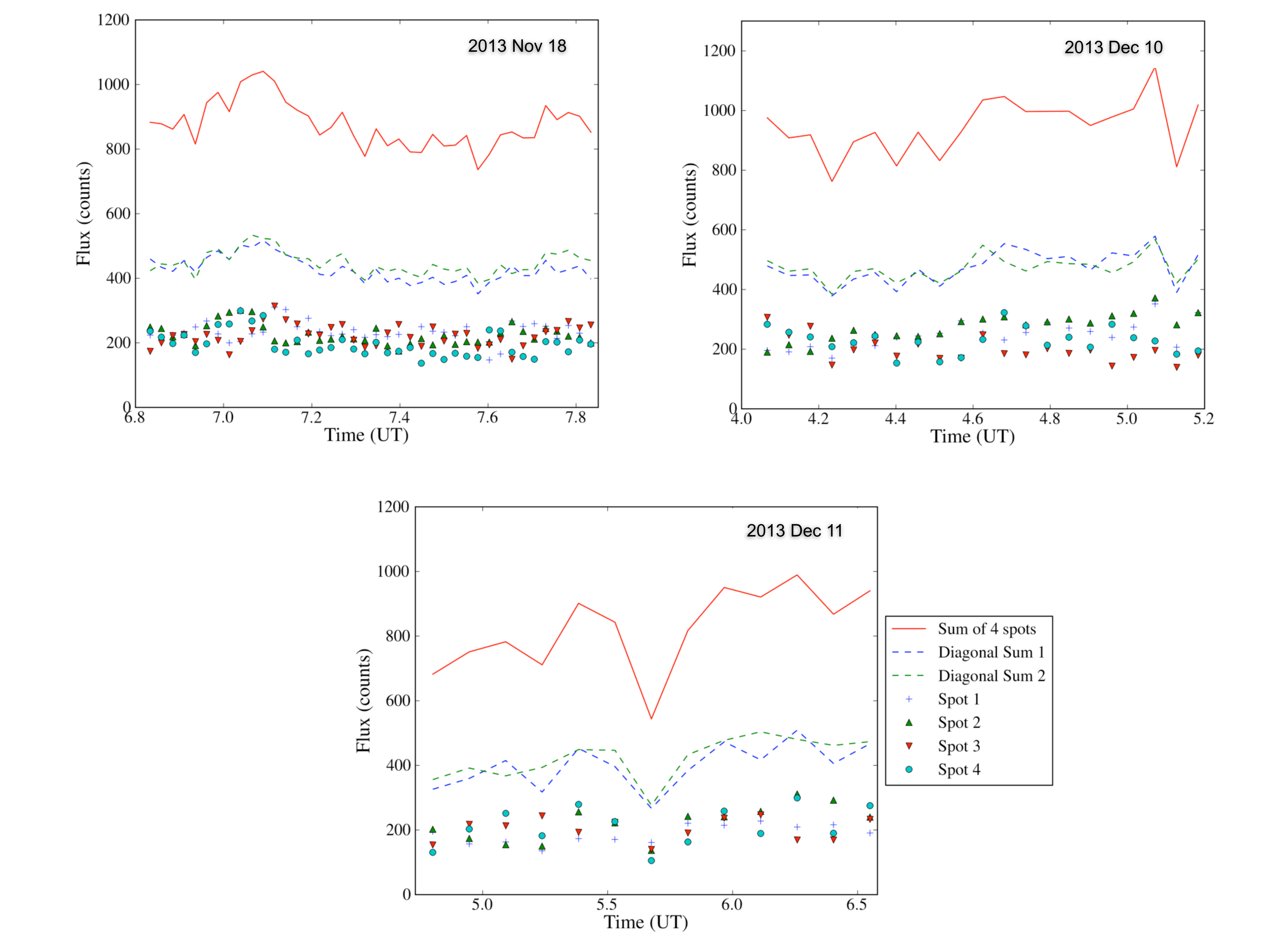}
   \end{center}
   \caption[example] 
%>>>> use \label inside caption to get Fig. number with \ref{}
   { \label{fig:bpmulti} GPI photometric monitoring of $\beta$~Pic. Plotted in the figure are the individual spot photometry in a 5~pixel aperture for a single wavelength slice. Also plotted are the sum of the two diagonal spots and all four spots represented by the dashed and solid lines. The sum of all four spots has the lowest relative scatter amongst the various combinations.}
   \end{figure} 
%-------------

\begin{figure}
   \begin{center}
   \includegraphics[height=12.cm]{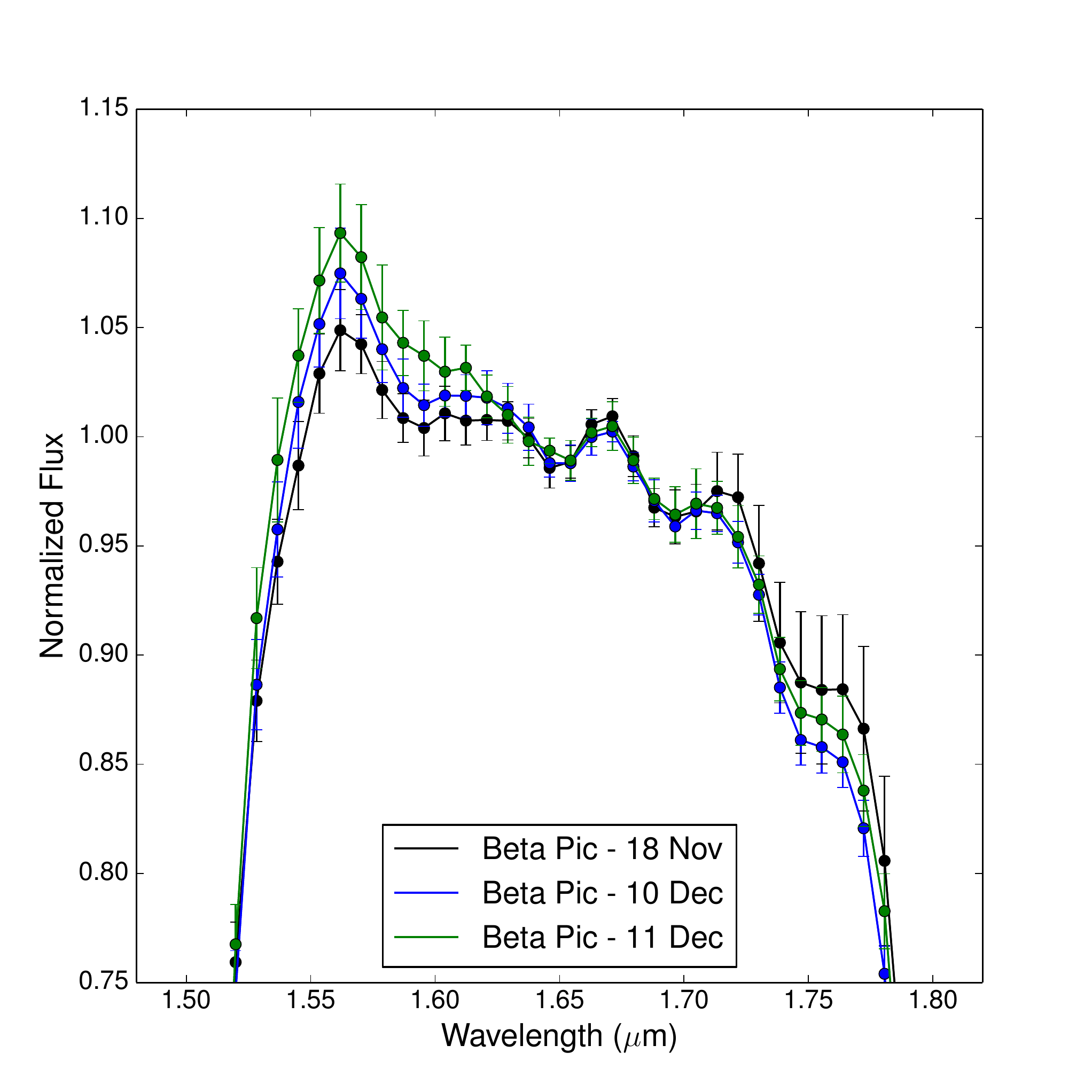}
   \end{center}
   \caption[example] 
%>>>> use \label inside caption to get Fig. number with \ref{}
   { \label{fig:bpmultispec} GPI spectrophotometric monitoring of $\beta$~Pic. Plotted in the figure is the spot spectrum taken over three epochs during the GPI commissioning run. The data are all normalized to the same region of the spectrum and the error bars are indicative of the scatter and the stability of the instrument over these short exposure times.}
   \end{figure}

Figure~\ref{fig:bpmulti} and \ref{fig:bpmultispec} show preliminary results on the long-term stability of the GPI spot data for $\beta$~Pic, used as a proxy for the stability of the integral field spectrograph (IFS). In Figure~\ref{fig:bpmulti} we have analyzed a single slice near the middle of the wavelength cube (1.662~$\mu$m), to estimate the stability of the spot photometry. The aperture photometry was performed using a 5-pixel radius aperture on unsharp-masked (9 px box median filtered) images. Plotted in the figure are the trends for each of the individual satellite spots, the sum of the diagonal spots and the sum of all four spots in the field. The sum of all four spots has the lowest scatter amongst the different trends plotted on the figure. The sum of spots is used for the measurement of spectrophotometric monitoring over shorter time durations in the next section. Figure~\ref{fig:bpmultispec} shows the same dataset as the one used for Figure~\ref{fig:bpmulti}, however it plots the spectral response of the sum of all four satellite spots for each of the spectral cubes observed over the three nights. The spectra have been all been normalized over the same spectral range to enable comparison of the scatter in the dataset, and the residual scatter over the three nights is $\sim$2\% indicating that the spot spectra are fairly stable and consistent across the three nights.

%-------------
%-------------

%\subsection{Spectral Mode}

\subsection{Short duration monitoring}
\label{sec:phot-spec}

\begin{figure}
   \begin{center}
   \includegraphics[height=7cm,trim=1cm 0cm 3.5cm 0cm]{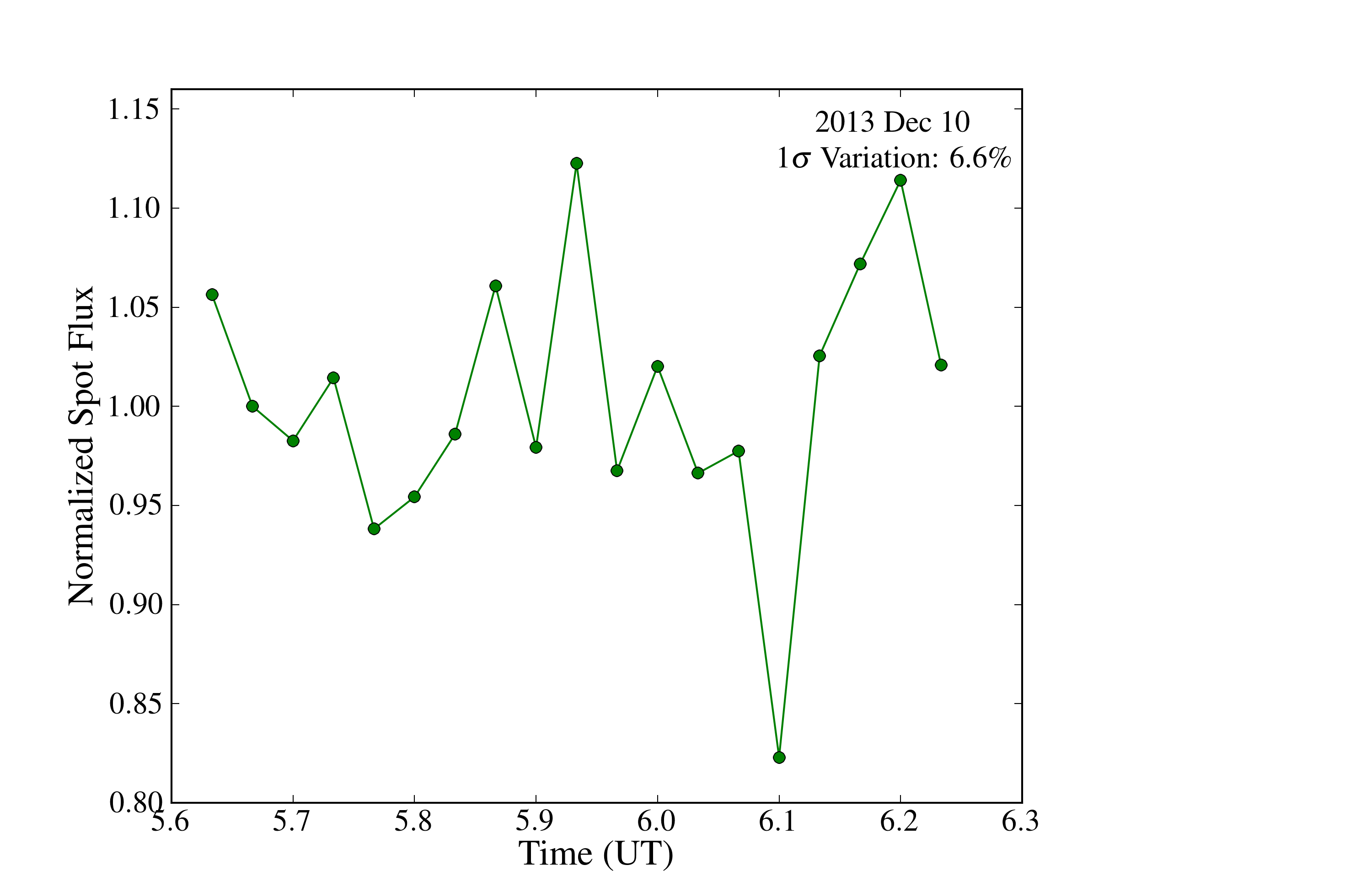}
   \includegraphics[height=7cm,trim=3cm 0cm 3.5cm 0cm]{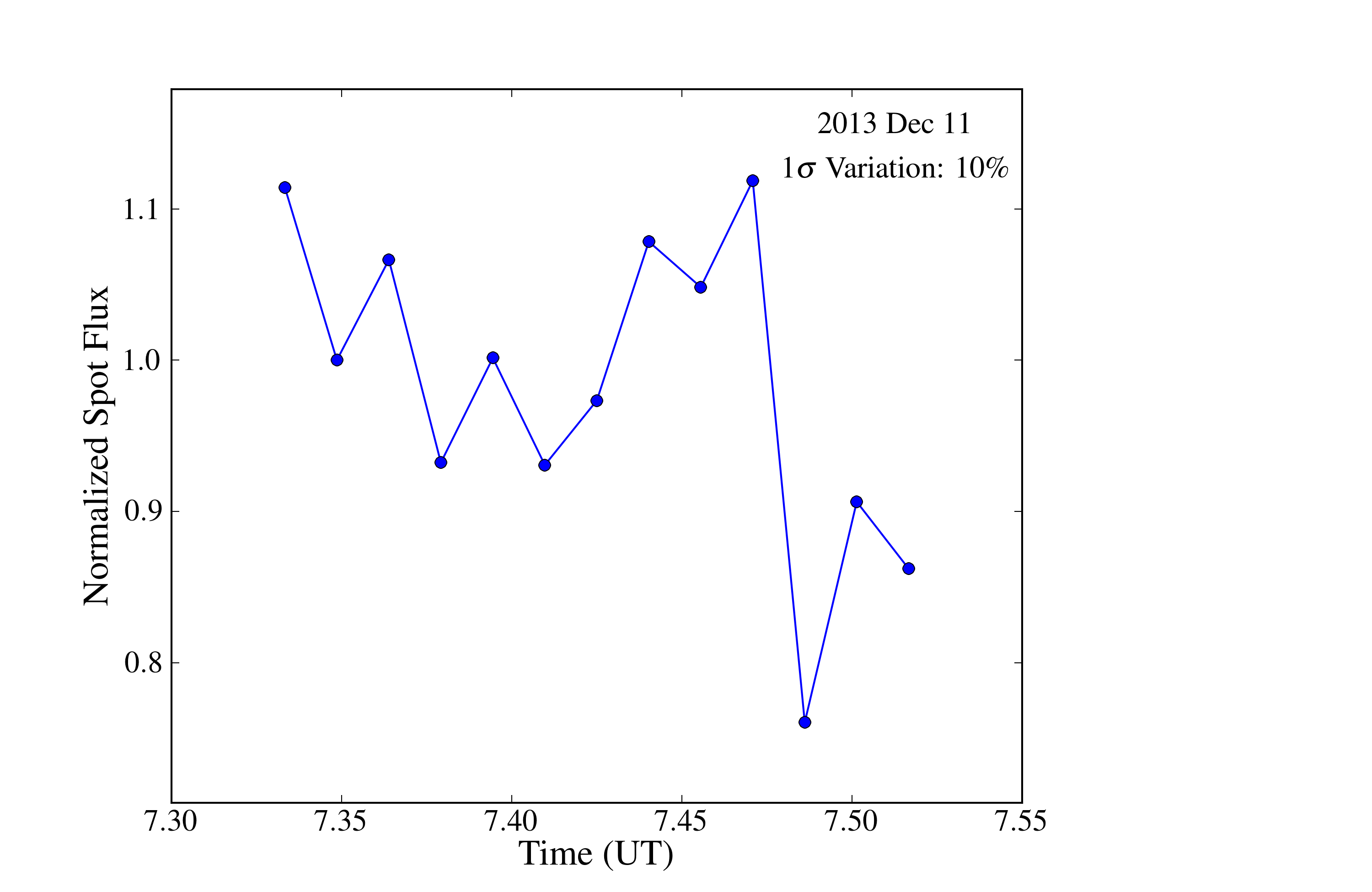}
   \end{center}
   \caption[example] 
   { \label{fig:bpicJ}  GPI photometric stability in the $J$-band. Measurements involve two different stars ($\beta$~Pic and HIP~47115) with short time-series taken on the 10 and 12 Dec 2013. Aperture photometry with a 5~px aperture was used to estimate the flux for all the data.}
   \end{figure}

For the $J$-band, two separate continuous data sets were identified, each of which span time periods of $\sim$15-25~min. These data were reduced using standard methods as described in Section 2, and then we measured aperture photometry of the spots using \texttt{DAOPHOT}. Since the data is collected over a short duration and taken at low airmass, we chose to not correct the photometry for airmass or seeing variations. For the $J$-band data we chose two separate targets, $\beta$~Pic and HIP~47115, which is of similar brightness as HR~8799.  Figure~\ref{fig:bpicJ} shows the sum of the spots (which was found to be the most stable choice) over the time duration. For the stack of 60s exposures, the $\beta$~Pic $J$-band data shows $\sim$6\% scatter over the duration of the data and for HIP~47115 it is $\sim$10\%. 

%-------------
%-------------

%-------------
   \begin{figure}
   \begin{center}
   \includegraphics[height=6.5cm]{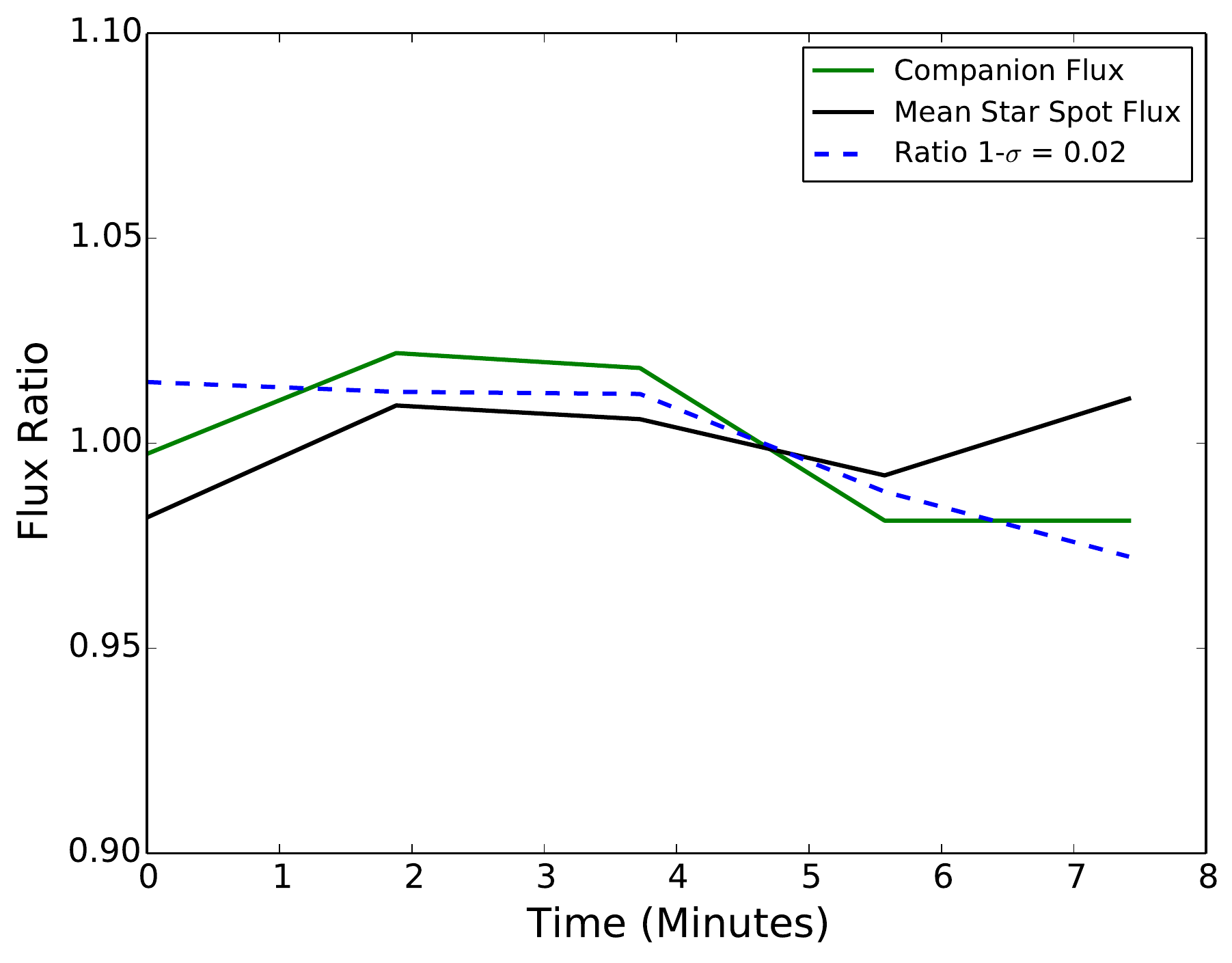}
   \includegraphics[height=6.5cm]{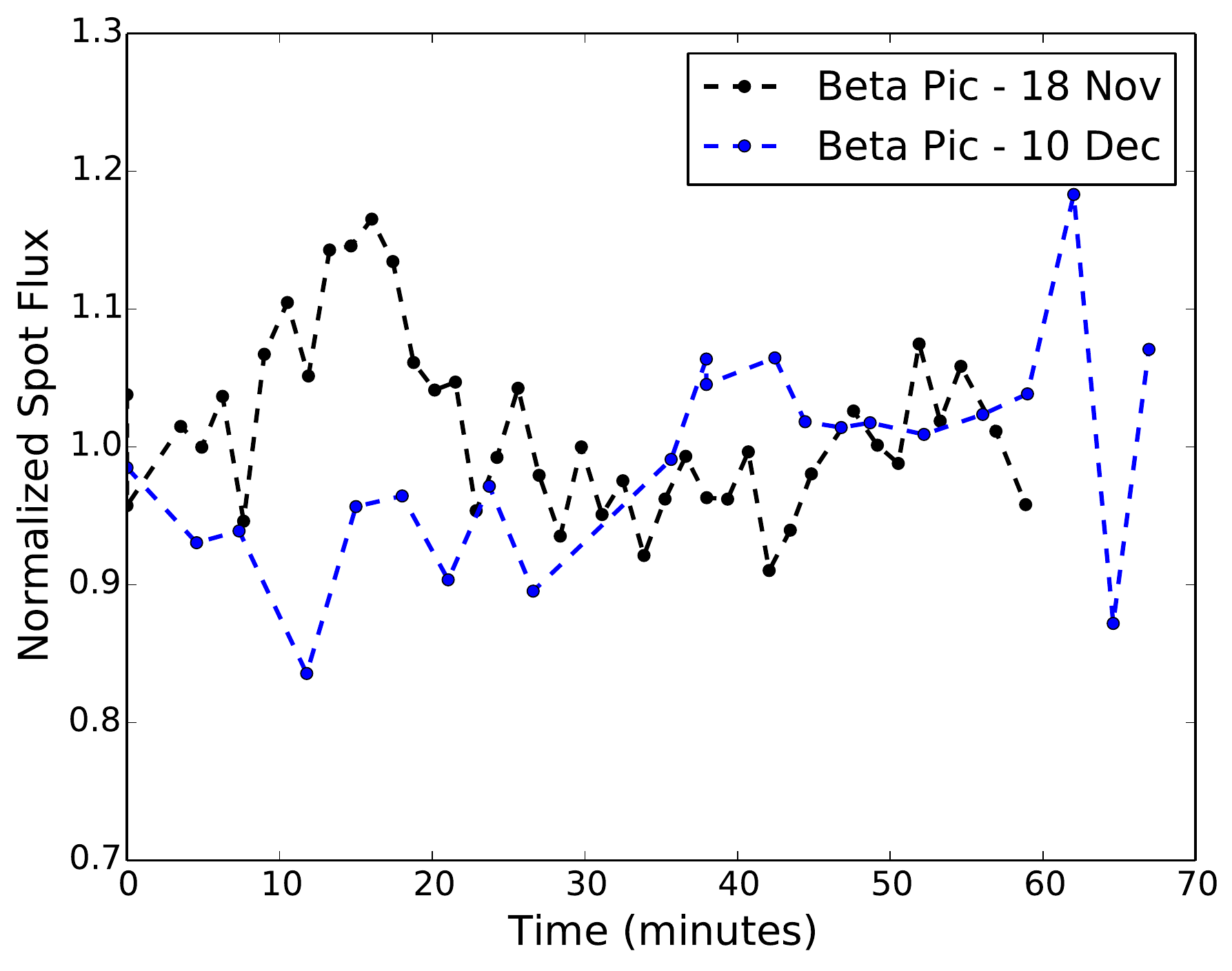}
   \end{center}
   \caption[example] 
   { \label{fig:8049H} GPI photometric stability in the $H$-band. Measurements involve using two different stars ($\beta$~Pic and HD~8049) with short and long time baseline observations. Aperture photometry with a 5~px aperture was used to estimate the flux for the entire data set.}
   \end{figure} 
%-------------

For the $H$-band, we utilized data taken on $\beta$~Pic and HD~8049 both of which have companions visible within the GPI field-of-view. Figure~\ref{fig:8049H} shows the variations in the data. The normalized spot flux has variations of $\sim7\%$ in the longer duration $\beta$~Pic A spot light curves and $\sim$2\% and $\sim$1\% variations in the spots and companion in the HD~8049 system respectively. %The time sequence only covers about 10~min which makes it difficult to quantify long-term stability of the photometry. This is a nice test object, since the white dwarf companion isn’t really expected to have astrophysical variations. The drawback is the shorter time of the sequence and the fainter spots which is a consequence of the faint primary star.

%%%%%%%%%%%%%%%%%%%%%%%%%%%%%%%%%%%%%%%%%%%%%%%%%%%%%%%%%%%%%%
\section{Future Work}
\label{sec:future}
Future work on improving satellite spot astrometric precision will proceed in three main areas. First, we will investigate the time-varying errors that are likely hindering the satellite spot astrometric precision. We will look at logs from the AO system to try to correlate the variations we see in the PSFs and the measured binary separation to anything physical. Next, we can improve the theoretical photon noise limit on relative centroiding within a datacube by combining satellite spots measured in all wavelength channels in a datacube to locate the occulted star. We will develop a least-squares fit to all the satellite spots to calculate a center, compensating for ADR or residual ADR if data is taken with the ADC. Having a model that connects the measurements from all 37 wavelength channels can improve robustness in locating the central star (e.g. insensitive to one bad satellite spot location) and help with images with low SNR satellite spots where the four satellite spots in a single wavelength slice are not precise enough. Lastly, we will continue searching for an astrometric calibrator suitable for GPI that will allow us to make an absolute calibration on satellite spot astrometry.

Future work on measuring the spectrophotometric stability of the GPI spots will require longer duration datasets, preferably over multiple hours in all the wavelengths. The length of the datasets we have now, less than $\sim40$~minutes in all wavelengths aside from the $H$-band, currently limits the measurements. Furthermore we need to measure the stability not just as a function of time but also across the different wavelengths. Some multi-epoch and multi-wavelength data exists for this work and further data will be collected over the final commissioning run to completely characterize the spot spectrophotometric stability. Additionally, we will develop techniques to measure and characterize satellite spot fluxes in polarimetry mode.

%%%%%%%%%%%%%%%%%%%%%%%%%%%%%%%%%%%%%%%%%%%%%%%%%%%%%%%%%%%%%%
\acknowledgments     %>>>> equivalent to \section*{ACKNOWLEDGMENTS}       
 
The Gemini Observatory is operated by the Association of Universities for Research in
Astronomy, Inc., under a cooperative agreement with the NSF on behalf of the Gemini
partnership: the National Science Foundation (United States), the National Research
Council (Canada), CONICYT (Chile), the Australian Research Council (Australia),
Minist\'erio da Ci\'encia, Tecnologia e Inova\c{c}\=ao (Brazil), and Ministerio de Ciencia,
Tecnolog\'ia e Innovaci\'on Productiva (Argentina). 
J.R.G., J.J.W., and P.K. thank support from NASA NNX11AD21G, NSF AST-0909188, and the University of California LFRP-118057. %James and Paul
This work is also supported in part by NASA grant APRA08-0117 and the STScI Director’s Discretionary Research Fund. %Anand
A.Z.G. is supported by the National Science Foundation Graduate Research Fellowship Program under Grant No. DGE-1232825 %Alex

%%%%%%%%%%%%%%%%%%%%%%%%%%%%%%%%%%%%%%%%%%%%%%%%%%%%%%%%%%%%%
%%%%% References %%%%%

\bibliography{report}   %>>>> bibliography data in report.bib
\bibliographystyle{spiebib}   %>>>> makes bibtex use spiebib.bst

\end{document}